\documentclass[twocolumn,amsmath,amssymb,longbibliography]{revtex4-2}

\usepackage[dvipdfmx]{graphicx}

\usepackage{bm}       
\usepackage{color}    
\usepackage{braket}   
\usepackage{mathdots} 
\usepackage{topcapt}  
\usepackage{booktabs} 
\usepackage[version=3]{mhchem} 

\usepackage[pdftex]{hyperref}
\hypersetup{
	colorlinks=true
}

\newif\iftitle
\titletrue

\newcommand  {\eqn}[1]{(\ref{eqn:#1})}
\renewcommand{\(}     {\left(}
\renewcommand{\)}     {\right)}
\renewcommand{\[}     {\left[}
\renewcommand{\]}     {\right]}
\renewcommand{\_}[1]  {_\mathrm{#1}}

\begin{document}

\title{
Ferromagnetic-electrodes-induced Hall effect in topological Dirac semimetals
}
\author{Koji Kobayashi$^1$}
\author{Kentaro Nomura$^{1,2}$}
\affiliation{$^1$Institute for Materials Research, Tohoku University, Sendai 980-8577, Japan}
\affiliation{$^2$Center for Spintronics Research Network, Tohoku University, Sendai 980-8577, Japan}

\begin{abstract}
 We propose an unconventional type of Hall effect in a topological Dirac semimetal with ferromagnetic electrodes.
 The topological Dirac semimetal itself has time-reversal symmetry, 
whereas attached ferromagnetic electrodes break it, causing the large Hall response.
 This induced Hall effect is a characteristic of the helical surface/edge states that arise in topological materials, such as topological Dirac semimetals or quantum spin Hall insulators. 
 We compute the Hall conductance/resistance and the Hall angle by using a lattice model with four-terminal geometry.
 For topological Dirac semimetals with four electrodes, the induced Hall effect occurs whether the current electrodes or the voltage electrodes are ferromagnetic.
 When the spins in electrodes are almost fully polarized, the Hall angle becomes as large as that of quantum Hall states or ideal magnetic Weyl semimetals.
 We show the robustness of the induced Hall effect against impurities and also discuss the spin injection
and spin decay problems.
 This Hall response can be used to detect whether the magnetizations of the two ferromagnetic electrodes are parallel or antiparallel.
\end{abstract}

\maketitle

\iftitle
\section{Introduction} \label{sec:intro}
\else
 \textit{Introduction.}
\fi

 The Hall effect occurs when a magnetic field is applied or when both magnetic ordering and spin-orbit coupling are present.
 The former refers to the ordinary Hall effect and the latter the anomalous Hall effect \cite{Nagaosa10anomalous}.
 The ordinary Hall effect results from Lorentz force in a magnetic field, 
while the anomalous Hall effect is known to be caused by spin-orbit coupled scattering (the extrinsic effect) or spin-orbit coupled band structure (the intrinsic effect).
 In any case, the time-reversal symmetry breaking in the bulk plays an essential role in the Hall effects.
 In this work, we propose an alternative way to obtain the Hall effect without breaking the time-reversal symmetry in the bulk;
the Hall effect can be induced by attaching the \textit{ferromagnetic} electrodes to \textit{nonmagnetic} topological Dirac semimetals (TDSs).

\begin{figure}[tbp]
 \centering
  \includegraphics[width=1\linewidth]{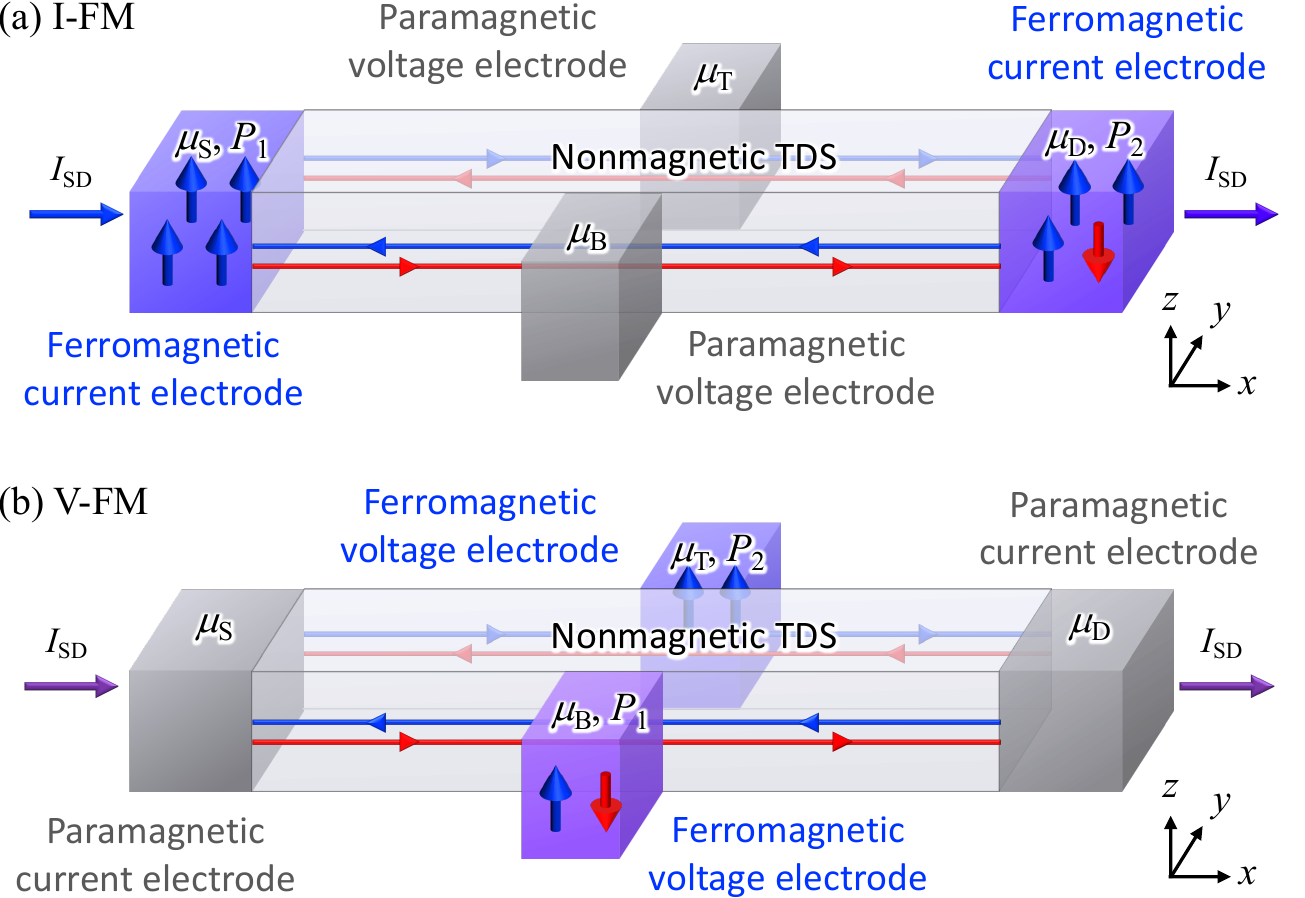}
 \vspace{-5mm}
\caption{
  Schematic figures of the four-terminal geometries:
  (a) the current electrodes are ferromagnetic while the voltage electrodes are paramagnetic (I-FM),
  and (b) the voltage electrodes are ferromagnetic while the current electrodes are paramagnetic (V-FM).
  The aspect ratio of the system is set to be $L_x\times L_y\times L_z = 6L\times L\times L$.
  The voltage electrodes of width $1$ are attached at the center of the top/bottom sides.
  The blue and red allows on the surface of TDS represent the current carrying states with up and down spins: the helical surface states.
}
\label{fig:system}
\end{figure}

 The TDS is a topological state of quantum matter
realized in \ce{Na3Bi} \cite{Wang12dirac, Liu14discovery}, \ce{Cd3As2} \cite{Wang13three, Neupane14observation}, 
CuMnAs \cite{Tang16dirac}, or Co$_{3-x}$Ni$_x$Sn$_2$S$_2$ \cite{Thakur20intrinsic,Shen20local},
for example.
 The TDS is characterized by a pair of band touching points (Dirac points) in the bulk and the surface helical Fermi arcs connecting the Dirac points \cite{Burkov16topological, Yan17topological, Armitage18weyl}.
 On the surface, spin-up electrons travel in one direction,
and spin-down electrons travel in the opposite direction (red and blue lines with arrows shown in Fig.~\ref{fig:system}).
 The TDS exhibits the intrinsic semi-quantized spin Hall effect
\cite{Morimoto14weyl,Burkov16z2,Taguchi20spin}
and is regarded as a three-dimensional analogue of the two-dimensional quantum spin Hall insulators \cite{Hasan10topological,Qi11topological}.
 The Hamiltonian of the TDS is time-reversal invariant,
and the net (charge) Hall effect cancels in the bulk TDS.

 The ferromagnetic-electrodes-induced Hall effect in the TDS is anticipated as follows.
 We first consider a geometry where the current electrodes (referred to as the source and drain electrodes) are spin-polarized [see Fig.~\ref{fig:system}(a)].
 Here the $z$ axis is set so that the pair of Dirac points is separated along the $k_z$ axis,
while the $x$ axis is the current flowing direction.
 The Hall response refers to an electric potential difference generated in the $y$ direction.
 The helical surface states appear on the surfaces parallel to the $z$ axis, 
which mediate the current from the source to drain electrodes.
 The spin-up electrons injected from the source electrode propagate only on one surface, i.e., the top surface, and 
enter the drain electrode.
 Spin-down electrons propagate on the other surface, i.e., the bottom surface.
 When the source and drain electrodes are spin-polarized and more spin-up electrons exist than spin-down electrons, 
more current flows on the top surface than on the bottom surface.
 As the top and bottom electrodes are in thermal equilibrium with electrons on the surface,
a finite chemical potential difference arises.
 Thus, the Hall response occurs even if the bulk remains time-reversal invariant.

 The Hall effect due to the injected spin-polarized current can also be explained by an analogy of the inverse spin-Hall effect,
which is a Hall effect induced by a pure spin current in the bulk.
 However, the ferromagnetic-electrodes-induced Hall effect can be realized even without net spin current in the bulk.
 The Hall effect occurs when the voltage electrodes are ferromagnetic
[see Fig.~\ref{fig:system}(b)].
 In this case, the Hall voltage arises because the helical surface states carry spin-up/down current on the top/bottom surface, respectively, and 
less electrons flow into the top/bottom electrodes if the spin-up/down state is the minority state, respectively.
 We found that the ferromagnetic-voltage-electrodes-induced Hall effect shows qualitatively the same property as the ferromagnetic-current-electrodes-induced Hall effect;
the Hall response is proportional to the polarization of the electrodes for parallel magnetizations and
vanishes for antiparallel magnetizations.
 Since the spin current is not injected,
the ferromagnetic-voltage-electrodes-induced Hall effect does not suffer from the conductance mismatch and spin decay problems.
 These ferromagnetic-electrodes-induced Hall effect enables us to measure the magnetization of the electrode by the Hall response and
is promising as a future spintronics material such as a magnetic memory device and spin MOSFET \cite{Zutic04spintronics, Sugahara04a}.

 The paper is organized as follows.
 In Sec.~II, we introduce a simple lattice model Hamiltonian for the description of TDSs.
 In Sec.~III~A, we numerically calculate transport properties and show that a Hall effect is induced in TDSs by ferromagnetic electrodes.
 In Sec.~III~B, we discuss the efficiency of the ferromagnetic-electrodes-induced Hall effect and the problem of spin injection.
 In Sec.~III~C, we propose that the ferromagnetic-electrodes-induced Hall effect causes the Hall magnetoresistance effect with diverging magnetoresistance ratio.
 In Sec.~IV, we introduce a normal metallic model with spin scatterers and show that the large ferromagnetic-electrodes-induced Hall effect is a feature of TDSs.
 The effect of spin diffusion and stability of the ferromagnetic-electrodes-induced Hall effect in TDSs is discussed in Sec.~IV~B.
 Then the paper is concluded in Sec.~V.

\iftitle\section{Model and Method} \label{sec:model}\fi
 We employ a simple cubic-lattice model
that describes the generic properties of TDSs around the Dirac point.
 The tight-binding Hamiltonian is \cite{Wang12dirac}
\begin{align}
 H &=  \sum_{\bf r}
         \[ {it \over 2} 
           \(
              \Ket{{\bf r}+{\bf e}_x}
               \tau_x \sigma_z
              \Bra{\bf r}
            + \Ket{{\bf r}+{\bf e}_y}
               \tau_y \sigma_0
              \Bra{\bf r}
           \)
           + \textrm{H.c.}
         \] \nonumber \\
   &+  \sum_{\bf r}
        \sum_{\mu=x,y,z}
         \[ \Ket{{\bf r}+{\bf e}_\mu}
           \(
              -{m_2 \over 2} \tau_z\sigma_0
           \)
           \Bra{\bf r}  + \textrm{H.c.}
         \]   \nonumber \\
   & + \sum_{\bf r} \Ket{\bf r}
        \(m_0 \tau_z\sigma_0
        \) \Bra{\bf r},
 \label{eqn:H_TDS}
\end{align}
where $\bm{\tau}$ and $\bm{\sigma}$ are Pauli matrices and correspond to orbital and spin degrees of freedom, respectively.
 Note that the model is spin $s_z$ conserved unless the spin scattering is induced (in Sec.~\ref{sec:extrinsic}).
 The topological property of an $s_z$-conserved TDS thin film is characterized by the spin Chern number \cite{Sheng06quantum},
which is equivalent to the Chern number for spin subbands and the number of the helical surface states on a side surface.
 The Chern number for a spin sector in this model can be obtained easily \cite{Yoshimura16comparative}.
 We set $m_0/m_2=2$ so that the Dirac points locate at $(0,0,\pm \pi/2)$ in the momentum space.
 Then, the spin Chern number $\mathcal{N}_s$ of the thin film is determined by the thickness $L_z$ as $\mathcal{N}_s=L_z/2$.
 We mention that the discussions in this paper is not limited to a specific TDS 
but apply to any system with helical edge states (i.e., with nonzero spin Chern number).
 In fact, the ferromagnetic-electrodes-induced Hall Effect is obtained also in quantum spin Hall systems
(see Appendix \ref{sec:2D}) or effective models with material parameters of \ce{Na3Bi} and \ce{Cd3As2} (see Appendix \ref{sec:material}).
 We set the hopping parameters $t=2$ and $m_2=1$,
and fix the Fermi energy at the Dirac point $E=0$ 
unless mentioned.

 The electrodes are semi-infinite one-dimensional ideal metallic wires 
 (where the hopping $t=2$ and the energy is located at the band center),
which couples to the TDS with the hopping parameter $t^\prime$.
 The ferromagnetic (FM) electrodes are realized by multiplying the weight for the up and down spin channels in the electrodes, $n_{\uparrow}$ and $n_{\downarrow}$, respectively.
 The magnetization (strength and angle) of the FM electrode is represented by the polarization, which is defined as 
 $P=\frac{n_{\uparrow}-n_{\downarrow}}{n_{\uparrow}+n_{\downarrow}}$.
 We ignore the exchange coupling in the TDS
for simplicity.
 Since the exchange splitting makes the TDS ferromagnetic \cite{Uchida19ferromagnetic}, 
one can regard the region of TDS with exchange coupling as a part of the ferromagnetic electrode.
 In a configuration for a magnetic sensor, 
the ferromagnetic electrodes may be separated by barrier (e.g., MgO) layers.
 These details of the electrodes are included in the coupling strength $t^\prime$ and polarization $P$ of the electrodes.

 We consider two types of four-terminal geometries with FM electrodes shown in Fig.~\ref{fig:system}:
FM \textit{current} electrodes (I-FM) and FM \textit{voltage} electrodes (V-FM).
 We calculate resistances (in units of $h/e^2$) and conductances (in units of $e^2/h$) based on the Landauer-B\"{u}ttiker formula.
 The source-drain resistance $R\_{SD}$ and the Hall resistance $R\_H$ are defined as
\begin{align}
 R\_{SD}= \frac{\mu\_S-\mu\_D}{e I\_{SD}}, \quad
 R\_{H} = \frac{\mu\_B - \mu\_T}{e I\_{SD}},
 \label{eqn:R}
\end{align}
and the source-drain conductance $G\_{SD}$ and the Hall conductance $G\_H$ are defined as
\begin{align}
 G\_{SD} &= \frac{1}{R\_{SD}}, \quad
 G\_{H} = \frac{R\_H}{R\_H^2+R\_{SD}^2},
 \label{eqn:G}
\end{align}
where
$I\_{SD}$ is the current between the source and drain electrodes, with $\mu\_S$, $\mu\_D$, $\mu\_T$ and $\mu\_B$ are the chemical potentials of the source, drain, top voltage, and bottom voltage electrodes, respectively.
 The chemical potentials are calculated by the recursive Green function method \cite{Ando91quantum,Datta97electronic}.

\iftitle
 \section{Ferromagnetic-electrodes-induced~Hall~effect}
\label{sec:result}\fi

\begin{figure}[tbp]
 \centering
  \includegraphics[width=1\linewidth]{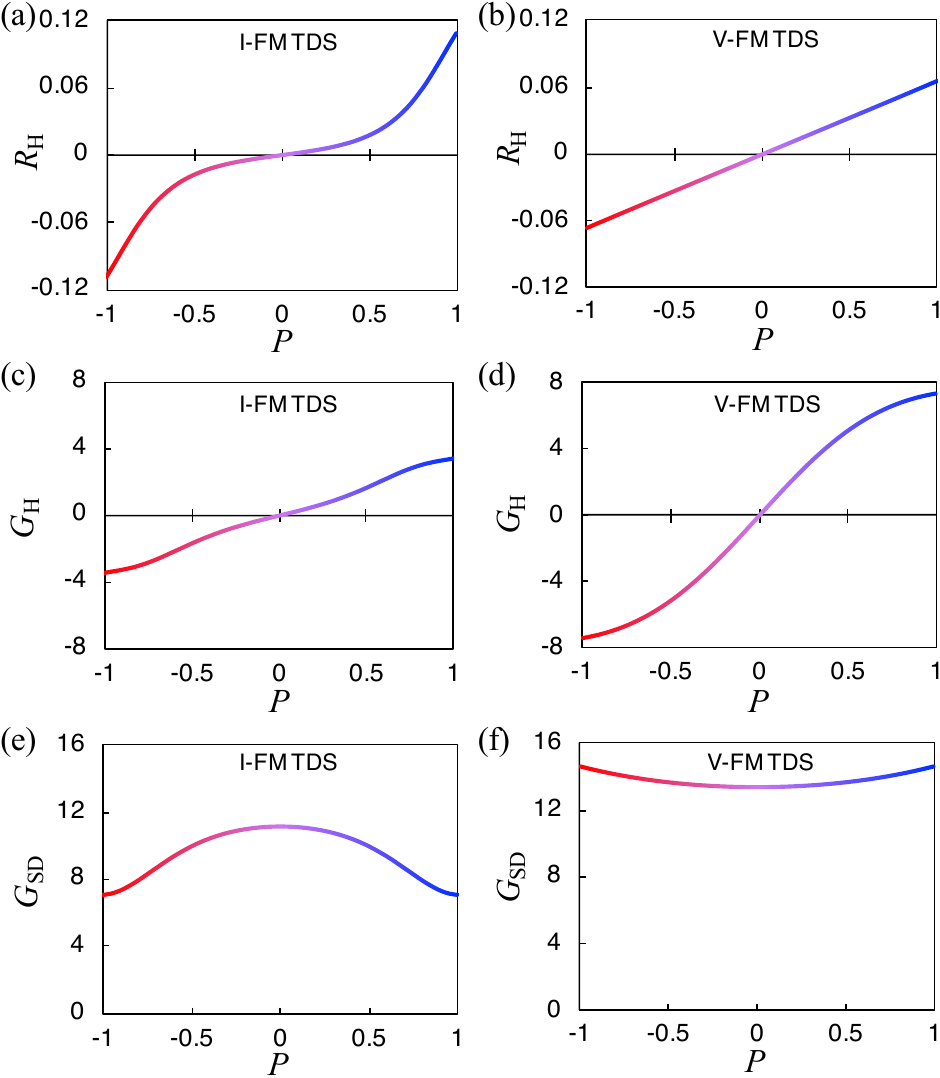}\\
  \includegraphics[width=1\linewidth]{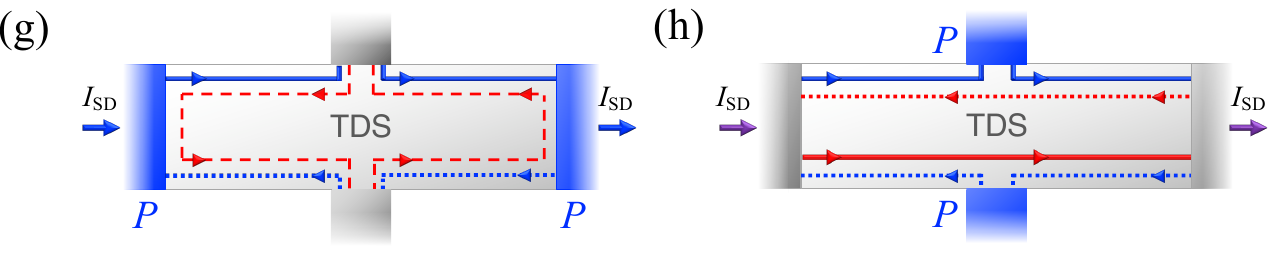}
 \vspace{-3mm}
\caption{
  (a)(b) Hall resistance $R\_H$, (c)(d) Hall conductance $G\_H$, and (e)(f) source-drain conductance $G\_{SD}$ as functions of the polarization of the FM electrodes $P=\frac{n_{\uparrow}-n_{\downarrow}}{n_{\uparrow}+n_{\downarrow}}$.
  The system size is $L=20$ 
  with the electrode coupling strength $t^\prime=t/2$.
  (g)(h) Schematic figure of the current flow in TDSs with $P=1$, i.e., half-metallic electrodes.
  Blue/red arrows represent the flow of spin-up/down current in the helical surface states.
  The left panels corresponds to the I-FM and the right panels to the V-FM TDSs.
}
\label{fig:GR}
\end{figure}

\iftitle
 \subsection{Polarization dependence}
\else
 \textit{Polarization dependence.}
\fi
 We first show that the Hall effect is induced by increasing the polarization of the electrodes.
 For simplicity, we consider the clean I-FM and V-FM TDSs with the same polarizations in the FM electrodes $P_1=P_2=P$.
 Figure~\ref{fig:GR} shows the calculated Hall resistance $R\_H$ and Hall conductance $G\_H$ as functions of the polarization of the electrodes $P$.
 For $P=0$, which means the density of states for up and down spins in the FM electrodes are equal (i.e., the electrodes are effectively paramagnetic),
the Hall resistance [Figs.~\ref{fig:GR}(a) and \ref{fig:GR}(b)] and conductance [Figs.~\ref{fig:GR}(c) and \ref{fig:GR}(d)] vanish.
 However, for a finite $P$, where the density of states for up and down spins are imbalanced,
a finite Hall effect is induced.
 The Hall response ($R\_H$ or $G\_H$) linearly rises, and the absolute value increases monotonically with increasing $|P|$.
 The Hall response is antisymmetric with respect to $P$ because the TDS is time-reversal symmetric (note that the time-reversal operation reverses spins).
 That is, the direction of the Hall response can be inverted by flipping the magnetizations of the FM electrodes.
 As one of the main results of this paper, we found that 
the transport properties of the TDS change drastically when the FM electrodes are attached and are sensitive to the magnetization of the electrodes; 
the ferromagnetic-electrodes-induced Hall effect arises even though the TDS itself is time-reversal symmetric.

 While the Hall conductance is enhanced with increasing polarization $|P|$ in both I-FM and V-FM TDSs [Figs.~\ref{fig:GR}(c) and \ref{fig:GR}(d)],
the longitudinal transport $G\_{SD}$ behaves differently for the I-FM and V-FM [Figs.~\ref{fig:GR}(e) and \ref{fig:GR}(f)].
 In the I-FM, $G\_{SD}$ decreases with increasing $|P|$
because the injected current is restricted to the spin-up states.
 In the V-FM, $G\_{SD}$ increases with increasing $|P|$
because the spin mixing in the electrodes is suppressed.
 To summarize, the TDS with FM electrodes shows a positive longitudinal magnetoresistance for the I-FM and a negative one for the V-FM
and shows the positive/negative ferromagnetic-electrodes-induced Hall effect at a positive/negative polarization.

\begin{figure}[tbp]
 \centering
  \includegraphics[width=1\linewidth]{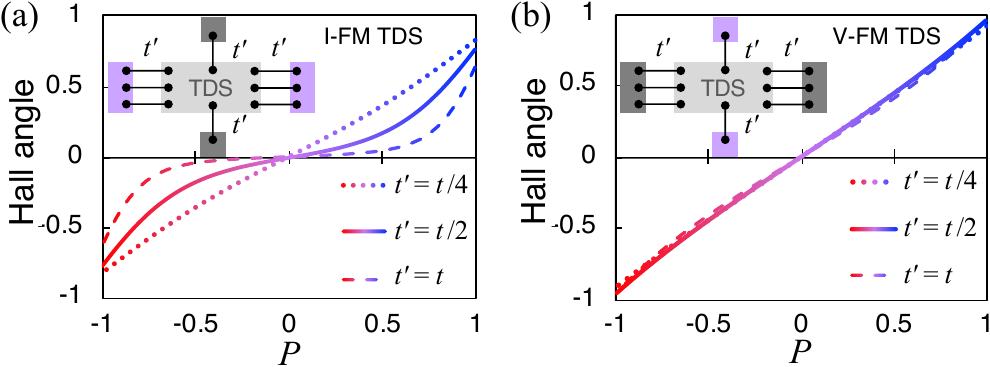}
 \vspace{-3mm}
\caption{
  Hall angle $R\_H/R\_{SD}$ as a function of the polarization of the current electrodes $P$ for the 
  (a) I-FM and (b) V-FM TDSs with $L=20$.
  The Hall angle for the I-FM TDS depends on the coupling strength of the electrodes $t^\prime=t/4$ (dotted), $t/2$ (solid), and $t$ (dashed),
  while that for the V-FM TDS does not.
}
\label{fig:HallAngle}
\end{figure}

 Next, we evaluate the strength of the ferromagnetic-electrodes-induced Hall effect.
 We use the ratio $R\_H/R\_{SD}$ as an indicator of the efficiency of the Hall effect and call it ``Hall angle'' 
(this corresponds to the ratio of the Hall and source-drain voltage $V\_H/V\_{SD}$, and 
is not equivalent to the ratio of the bulk Hall and longitudinal conductivities $\sigma\_H/\sigma_{xx}$).
 By focusing on the Hall angle, we can exclude the unessential dependence on the system size and aspect ratio.
 The Hall angle takes the maximum value $|R\_H/R\_{SD}|=1$ for the ideal quantum (anomalous) Hall effect,
while it is much smaller (typically, $|R\_H/R\_{SD}|\simeq 0.01$) for extrinsic Hall effects in conventional materials (we discuss later, in Sec.~\ref{sec:extrinsic}).
 The evaluated Hall angle of the ferromagnetic-electrodes-induced Hall effect
is shown in Fig.~\ref{fig:HallAngle}.
 For a half-metallic electrode $|P|\simeq 1$, the Hall angle 
is close to $1$, i.e., as large as the quantum (anomalous) Hall effect, 
in both the I-FM and V-FM TDSs.

\iftitle 
 \subsection{Spin injection}
\else
 \textit{Spin injection.}
\fi
\label{sec:CM}

  The Hall angle in I-FM TDSs [Fig.~\ref{fig:HallAngle}(a)] depends on the coupling parameter $t^\prime$ between the electrodes and TDS;
the Hall angle becomes large for a small $t^\prime$.
 This tells us the condition of the electrodes for obtaining a large ferromagnetic-electrodes-induced Hall effect
because the parameter $t^\prime$ involves the information of the electrode and interface
(e.g., the density of states in the electrode, lattice mismatch, and contact resistance).
 When the FM electrodes are fully polarized (half-metallic),
we obtain a large Hall angle irrespective of details of the electrodes.
 When the FM electrodes are partially polarized,
we can achieve a large Hall angle by using the FM electrodes with small $t^\prime$, 
say, with large contact resistance.

 The $t^\prime$ dependence of the Hall angle can be understood as a problem of the spin injection from metallic electrodes to semiconductors,
which is known as the conductance mismatch or impedance mismatch problem \cite{Zutic04spintronics}.
 In our model, we have three parameters relevant to the spin injection: the coupling $t^\prime$ and system size $L$, in addition to $P$.
 The coupling $t^\prime$ is related to the contact resistance.
 The number of the current-carrying states in a metallic electrode
is proportional to the cross section ($\sim L^2$),
while that in the TDS (the helical surface states) is proportional to the length along the direction of the Dirac nodes separation ($z$ axis), i.e., the thickness ($\sim L$).
 In order to understand the mismatch problem,
we have to investigate the dependence of the Hall angle on $t^\prime$ and $L$ independently.

\begin{figure}[tbp]
 \centering
  \includegraphics[width=1\linewidth]{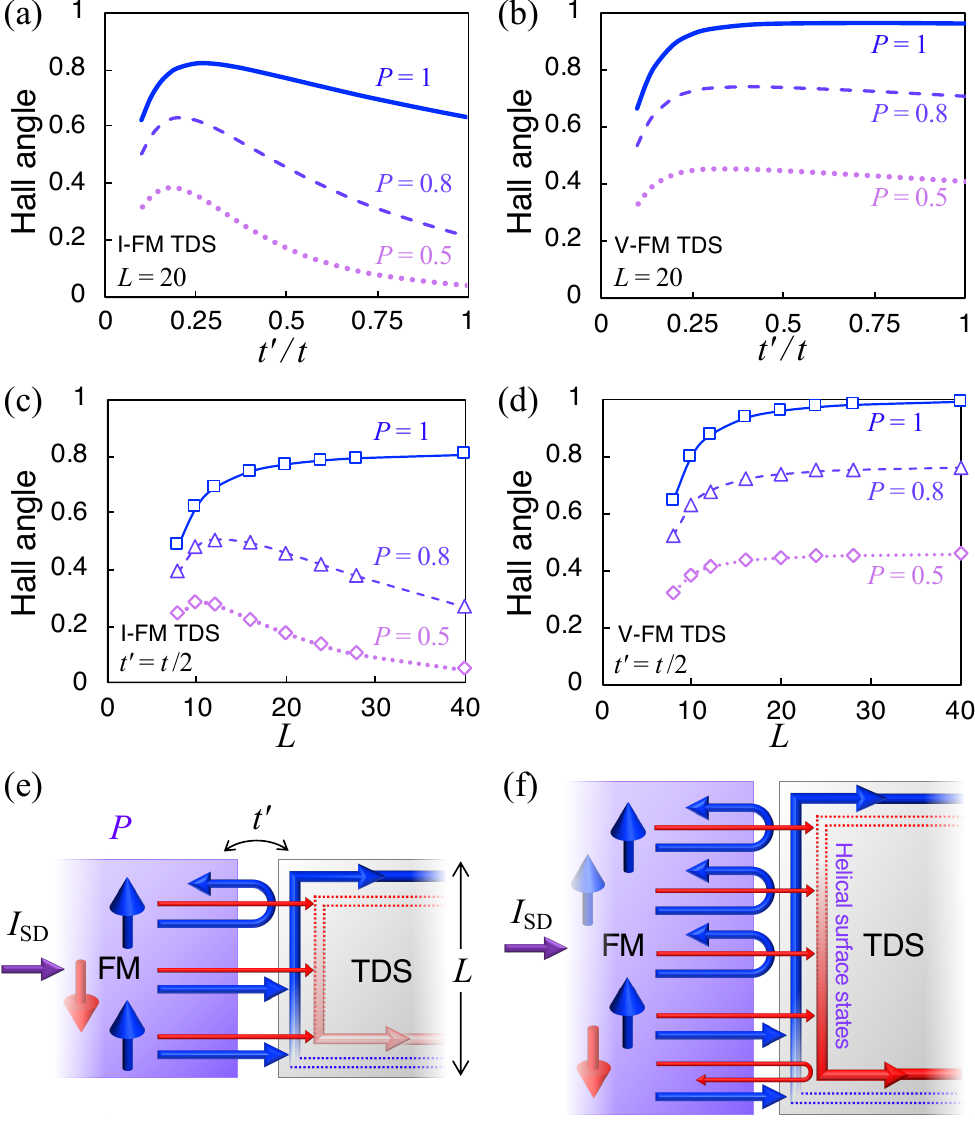}
 \vspace{-3mm}
\caption{
  Hall angle $R\_H/R\_{SD}$ as a function 
  (a)(b) of the coupling between the sample and electrodes $t^\prime$ with $L=20$, and
  (c)(d) of the system size $L$ with $t^\prime=t/2$,
  for $P=1$ (solid), $0.8$ (dashed), and $P=0.5$ (dotted),
  in the (a)(c) I-FM and (b)(d) V-FM TDSs.
  Schematic image of spin injection into the helical surface states 
  for (e) a weak $t^\prime$ and a small $L$, and
  for (f) a sufficiently strong $t^\prime$ or a large $L$.
}
\label{fig:CM}
\end{figure}

 First, we show the $t^\prime$ dependence of the Hall angle in Figs.~\ref{fig:CM}(a) and \ref{fig:CM}(b).
 In the I-FM TDS [Fig.~\ref{fig:CM}(a)], the Hall angle 
has a maximum at a small $t^\prime$ ($\simeq t/4$) 
and decreases with increasing $t^\prime$.
 In contrast, in the V-FM TDS, the Hall angle [Fig.~\ref{fig:CM}(b)] is not so sensitive to $t^\prime$ as in the I-FM TDS.
 Next, we show the $L$ dependence of the Hall angle in Figs.~\ref{fig:CM}(c) and \ref{fig:CM}(d).
 The Hall angle in the I-FM TDS [Fig.~\ref{fig:CM}(c)] decays exponentially with increasing system size $L$,
while that in the V-FM TDS [Fig.~\ref{fig:CM}(d)] increases and saturates for a large $L$.
 Here we note that the Hall angle depends mainly on the width $L_y$, 
though we have kept the aspect ratio for consistency in Fig.~\ref{fig:CM}.
 The mechanism of the decay of the Hall angle in the I-FM TDS can be explained as follows.
 When the source electrode is highly polarized ($P\simeq 1$),
the spin-up current
is injected into the TDS more than the spin-down current,
and hence the Hall effect arises [see Fig.~\ref{fig:CM}(e)].
 Here, since the number of the current-carrying states (the helical edge states) in the TDS
is much smaller than that in the metallic FM electrode,
the spin-up current cannot be injected exceeding the capacity of the helical edge states.
 On the other hand, 
the spin-down current can be accumulatively injected into the TDS
as the helical surface states run along the boundary [see Fig.~\ref{fig:CM}(f)].
 Therefore, when the length of the interface along the helical surface states 
$L_y$
is long enough or the coupling 
$t^\prime$
is strong enough,
the spin-down current becomes as large as the spin-up current.
 Hence the Hall angle is small for a large $t^\prime$ or a large $L$ in the I-FM TDS.

 We found a practical difference between the ferromagnetic-electrodes-induced Hall effects in the I-FM and V-FM TDSs in terms of the spin injection.
 The I-FM TDS tends to suffer from the spin injection
since the current electrodes are essentially metallic.
 In contrast, the V-FM TDS is relatively insensitive to the spin injection problem,
because we can make the voltage electrodes arbitrarily narrow or weak coupling.
 Therefore, even when the spin injection problem arises in actual materials,
we may overcome the problem by employing the V-FM geometry.
 The upper bound of the Hall angle $R\_H/R\_{SD}$ at a certain $P$ can be estimated from the extrapolation of the curves in Fig.~\ref{fig:CM}.
 We found, in both the I-FM and V-FM TDSs, that the upper bound of the Hall angle of the ferromagnetic-electrodes-induced Hall effect will be $P$ (i.e., $R\_H/R\_{SD}=P$ if we ignore the spin injection problem).

\iftitle
 \subsection{Hall magnetoresistance}
\else
 \textit{Hall magnetoresistance.}
\fi
 The bulk Hall effect with a large Hall angle can be achieved in magnetic Weyl semimetals (WSMs).
 The magnetic WSM \cite{Wan11topological, Yang11quantum, Burkov11weyl, Hosur13recent, Wang16time, Jin17ferromagnetic, Burkov16topological, Yan17topological, Armitage18weyl} is a time-reversal-broken topological state
and is one of the promising systems for next-generation spintronics devices
due to its fancy transport properties (e.g., huge longitudinal magnetoresistance effects \cite{Ominato17anisotropic, Kobayashi18helicity, Kobayashi19robust}).
 The magnetic WSM state is proposed in
magnetic kagome layered materials (Fe$_3$Sn$_2$ \cite{Ye18massive,Yin18giant}, Co$_3$Sn$_2$S$_2$ \cite{Liu18giant, Yin19negative, Ozawa19two, Liu19magnetic, Muechler20emerging, Shen2033, Tanaka20topological}, and Mn$_3$Sn \cite{Nakatsuji15large, Yang17topological, Ito17anomalous}), and
Heusler compounds (Ti$_2$MnAl \cite{Shi18prediction}, Co$_2$MnGa \cite{Sakai18giant, Guin19anomalous}, and Co$_2$MnAl \cite{Li20giant}).
 Especially, the Hall angle is reported as
 $0.3$ \cite{Shen2033,Tanaka20topological} for Co$_3$Sn$_2$S$_2$, $0.1$ \cite{Sakai18giant} for Co$_2$MnGa, and $0.2$ \cite{Li20giant} for Co$_2$MnAl.

 In comparison with the bulk anomalous Hall effect in magnetic WSMs,
the ferromagnetic-electrodes-induced Hall effect in TDSs has an advantage in controllability.
 That is, the strength and on/off of the ferromagnetic-electrodes-induced Hall effect can be controlled externally without modifying the TDS state.
 Changing the magnetization of FM electrodes will be much easier
since there is a broader range of candidates for the FM electrodes than the topological materials.
 Furthermore, since the FM electrodes are spatially separated,
we can change the magnetizations of the FM electrodes independently.
 This provides more application potentiality to the ferromagnetic-electrodes-induced Hall effect.

\begin{figure}[tbp]
 \centering
  \includegraphics[width=1.\linewidth]{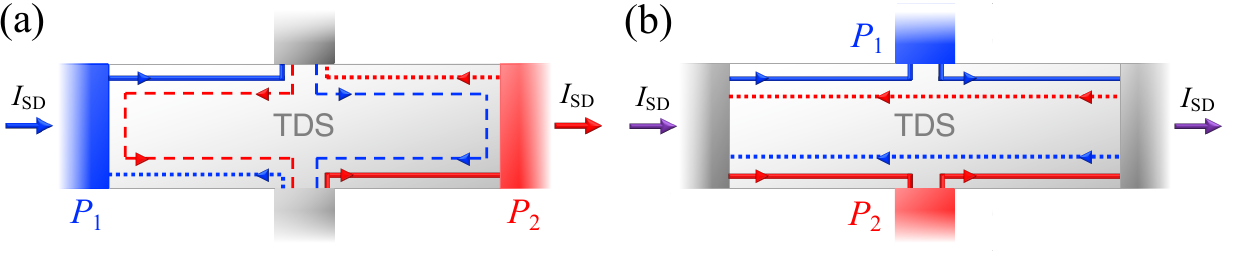}
 \vspace{-5mm}
\caption{
  Schematic figure of the spin-dependent current flow in (a) I-FM and (b) V-FM TDSs with antiparallelly magnetized FM electrodes. 
  Blue and red electrodes represent the polarization $P=1$ and $P=-1$, respectively.
  Blue/red arrows represent the flow of spin-up/down current in the helical surface states.
}
\label{fig:currentAP}
\end{figure}

\begin{figure}[tbp]
 \centering
  \includegraphics[width=1.0\linewidth]{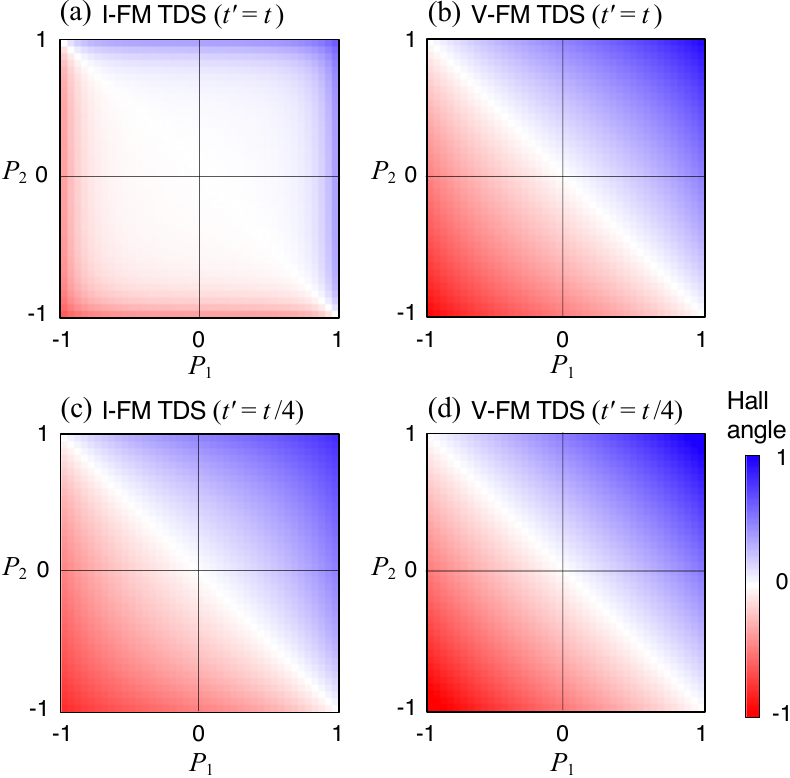}
 \vspace{-5mm}
\caption{
  Color map of the Hall angle $\frac{R\_H(P_1,P_2)}{R\_{SD}(P_1,P_2)}$
  with 
  (a)(b) a large electrodes coupling $t^\prime=t$ and 
  (c)(d) a small electrodes coupling $t^\prime=t/4$.
  Left/right panels are for the clean I-FM/V-FM TDS.
  The horizontal and vertical axes correspond to the polarization of the pair of FM electrodes.
  The system size is $L=20$.
}
\label{fig:PPmap}
\end{figure}

 We then show a feature of the ferromagnetic-electrodes-induced Hall effect that will be useful for the device applications: a huge Hall magnetoresistance effect.
 The I-FM configuration with parallel or antiparallel magnetizations corresponds to the CPP-GMR device \cite{Zutic04spintronics} with the barrier layer of TDS,
and we can achieve the longitudinal magnetoresistance effect.
 The longitudinal magnetoresistance in TDSs may be useful due to a long spin diffusion length (we discuss later, Sec.~\ref{sec:diffusion}).
 Meanwhile, the Hall magnetoresistance in TDSs may give an extraordinarily large magnetoresistance ratio.
 Because the sign of the ferromagnetic-electrodes-induced Hall effect corresponds to the direction (up/down) of the magnetizations in FM electrodes,
the effect cancels when the magnetizations are antiparallel (see Fig.~\ref{fig:currentAP}).
 This means that we can switch on/off the Hall resistance by flipping the magnetization of the electrodes.
 The difference of the Hall angle by changing the polarizations can be found in the Hall angle map shown in Fig.~\ref{fig:PPmap}.
 The diagonal line with $P_1=P_2$ shows the occurrence of a large ferromagnetic-electrodes-induced Hall effect when the magnetizations are parallel and high,
as shown above (Fig.~\ref{fig:HallAngle}).
 On the other hand, on the line $P_1=-P_2$ where the magnetizations are antiparallel, the Hall angle is always zero.
 Therefore, by changing the magnetization of the electrodes from antiparallel to parallel,
we can obtain a large Hall magnetoresistance effect.
 Note that the Hall magnetoresistance ratio, which is defined as $\frac{R\_{H}(P, P)-R\_{H}(P,-P)}{R\_{H}(P, -P)}$ or $\frac{R\_{H}(P_1, P_2)-R\_{H}(0,0)}{R\_{H}(0,0)}$, for example, always diverges theoretically.
 In practice, the Hall magnetoresistance ratio will be finite 
since the Hall resistance $R\_{H}(P_1, P_2)$ in the denominator is not exactly zero due to the asymmetry in the FM electrodes (small deviation of the polarization $P_1\neq -P_2$, size, or position) and the spin scattering (see Sec.~\ref{sec:extrinsic}).
 Thus, we expect a large Hall magnetoresistance ratio
unless the spin injection problem is crucial [Fig.~\ref{fig:PPmap}(a)] or $P_1,P_2\simeq 0$.
 We also expect the qualitative behavior of the Hall magnetoresistance is insensitive to the details
because the Hall angle maps are almost the same for the I-FM TDS with a large contact resistance [$t^\prime=t/4$, Fig.~\ref{fig:PPmap}(c)] and the V-FM TDSs [Figs.~\ref{fig:PPmap}(b) and \ref{fig:PPmap}(d)].

 When we fix the polarization of one of the FM electrodes $P_2$ to be zero (the vertical lines on $P_1=0$ in Fig.~\ref{fig:PPmap}),
the Hall angle becomes just one half of the value for $P_1=P_2=P$ (shown in Fig.~\ref{fig:HallAngle}).
 Specifically, the Hall angle is almost proportional to $P_1$,
and can be as large as $R\_H/R\_{SD}=0.5$ for $P_1=1$ or $R\_H/R\_{SD}=-0.5$ for $P_1=-1$.
 This means that we can detect the polarization (up/down and magnitude) of a sample from the Hall angle,
by preparing the four-terminal device with one of the four electrodes is FM (sample), and the others are paramagnetic.
 Although the FM sample can be either the current or voltage electrode,
the voltage electrode will be more efficient because of the lack of the spin injection problem, as discussed in Sec.~\ref{sec:CM}.

\begin{figure}[tbp]
 \centering
  \includegraphics[width=1.\linewidth]{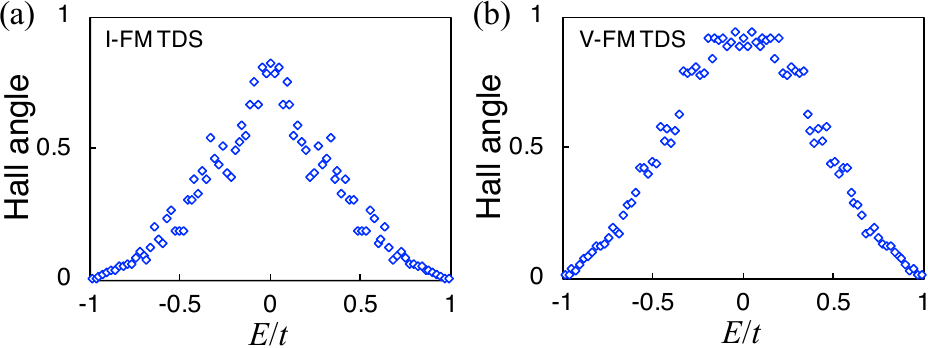}
 \vspace{-5mm}
\caption{
  Hall angle as a function of Fermi energy $E/t$ in (a) I-FM and (b) V-FM TDSs.
  The electrodes are half-metallic $P=1$ with coupling $t^\prime=t/4$, and the size $L=20$.
}
\label{fig:Edep}
\end{figure}

\iftitle
 \subsection{Energy dependence}
\else
 \textit{Energy dependence.}
\fi
 We here show that a large Hall angle obtained in a broad range of energy.
 The density of states $\rho$ of a bulk TDS behaves as $\rho\sim E^2$ around the Dirac point (in our model, this behavior holds up to $|E|\simeq t$).
 Since the density of states (and longitudinal conductivity) vanishes only at the Dirac point $E=0$,
one should be careful whether or not the obtained results are the singularity of $E=0$.
 The energy dependence of the Hall angle is shown in Fig.~\ref{fig:Edep}.
 In both I-FM and V-FM geometries, the Hall angle is large around the Dirac point 
and finite in the range $|E|<t$ where the Dirac cone (i.e., the quadratic behavior of $\rho$) arises.
 While the Hall angle shows fluctuations due to the discretized energy levels in finite-size systems,
the Hall angle in V-FM TDS [Fig.~\ref{fig:Edep}(b)] is relatively stable against the energy deviation.

\iftitle
 \section{Spin scattering}\label{sec:extrinsic}
\else
 \textit{Spin scattering.}
\fi

 Next, we discuss the effect of spin scattering.
 The spin scattering induces two effects:
the extrinsic Hall effect and the spin diffusion.
 In order to understand the extrinsic Hall effect,
we introduce a model for the normal metal (NM) on the cubic lattice,
\begin{align}
 H &=  \sum_{\bf r}
        \sum_{\mu=x,y,z}
         \[ \Ket{{\bf r}+{\bf e}_\mu}
           \(
              -{t \over 2} \sigma_0
           \)
           \Bra{\bf r}  + \textrm{H.c.}
         \], 
 \label{eqn:H_NM}
\end{align}
with the same hopping parameter as the TDS model ($t=2$).
 We fix the Fermi energy at the band center.

 We introduce spin scatterers $U_s \propto W_s \(\bm{s} \times \bm{p}\)\cdot \nabla V$ \cite{Nikolic07extrinsically}, as
\begin{align}
 U_s &=  \sum_{\bf r} \Ket{\bf r}
          \sum_{\mu,\nu,\gamma=x,y,z} \sum_{m,n=\pm 1}
          (- i W_s) \varepsilon_{\mu\nu\gamma} m n s_\gamma \nonumber \\
     &  \times \[   V({\bf r} + m{\bf e}_\mu) - V({\bf r} + n{\bf e}_\nu) 
          \] \Bra{{\bf r} + m{\bf e}_\mu + n{\bf e}_\nu},
 \label{eqn:H_s}
\end{align}
with the impurity potential $V({\bf r})$.
 For simplicity, we consider short-ranged impurities with height $1$, 
which are distributed randomly on the lattice sites with the density $\rho$.
 The spin operators $s_{x,z}=\tau_0 s_{x,z}$ and $s_{y}=\tau_z s_{y}$ for the TDS, and $s_i=\sigma_i$ for the NM model.
 We also introduce the impurity potential term
\begin{align}
 U_0 &=  \sum_{\bf r} \Ket{\bf r}
          \[ V({\bf r}) - \bar{V} \] s_0
          \Bra{{\bf r}},
 \label{eqn:H_p}
\end{align}
where the averaged potential $\bar{V}$ is substituted so as to keep the energy of the Dirac point ($E=0$) unchanged.

\begin{figure}[tbp]
 \centering
  \includegraphics[width=1\linewidth]{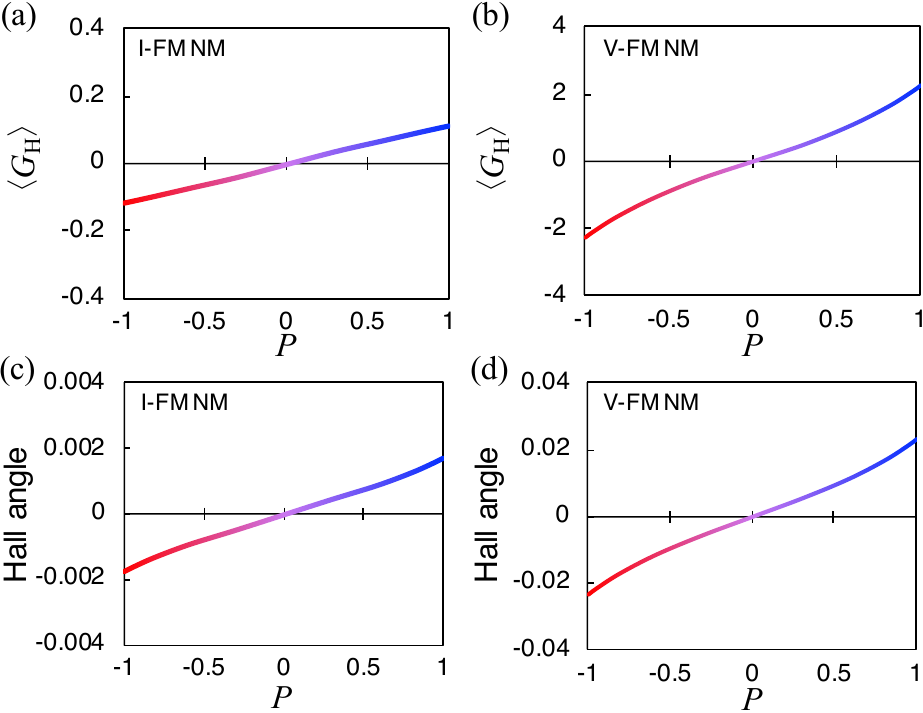}
 \vspace{-4mm}
\caption{
  (a)(b) Averaged Hall conductance $\braket{G\_H}$ and (c)(d) Hall angle $\braket{R\_H}/\braket{R\_{SD}}$
  in normal metals (NMs).
  Left and right panels are for the I-FM and V-FM NM geometries, respectively.
  The horizontal axis $P$ is the polarization of FM electrodes.
  Spin scatterers of $W_s=0.5$ and $\rho=5\%$ are induced, and
  the system size $L=20$ and coupling $t^\prime=t/2$.
  Averaged over 20,000 impurity realizations and the error bars are less than the width of the lines.
}
\label{fig:VFM_NM}
\end{figure}

\iftitle
 \subsection{Extrinsic Hall effect}
\else
 \textit{Extrinsic Hall effect.}
\fi
 We show that the ferromagnetic-electrodes-induced Hall effect can also be raised by the extrinsic mechanism, 
but its strength is much weaker than the intrinsic one in TDSs.
 In general, the origin of the anomalous Hall effect can be divided into two types: intrinsic and extrinsic.
 Typically, the former arises in systems with spin-orbit coupled bands, such as topological systems, and the latter comes from spin scattering by impurities.
 By introducing the spin scatterers into a FM metal, an extrinsic anomalous Hall effect occurs \cite{Nagaosa10anomalous}.
 In the same way, by introducing the spin scatterers into a \textit{nonmagnetic} metal with FM electrodes, 
an extrinsic ferromagnetic-electrodes-induced Hall effect arises (see Fig.~\ref{fig:VFM_NM}).
 Although the absolute value of the Hall conductance can be comparable to that in TDSs,
the Hall angle is a few percent at most.
 While the spin scattering generates the ferromagnetic-electrodes-induced Hall effect, 
it also breaks the spin conservation and destroys the ferromagnetic-electrodes-induced Hall effect.
 Thus, there are maxima in $G\_H$ and $R\_H$ as functions of the spin-scattering parameters $W_s$ and $\rho$.
 Even though we have tuned $W_s$ and $\rho$ so as to obtain the Hall effect efficiently (i.e., $W_s=0.5$ and $\rho=5\%$),
the Hall angle is still significantly smaller than that from the intrinsic ferromagnetic-electrodes-induced Hall effect in TDSs.
 Therefore, we can say that the ferromagnetic-electrodes-induced Hall effect with a large Hall angle is a feature of TDSs.
 This also implies that the extrinsic ferromagnetic-electrodes-induced Hall effect in TDSs, which have a smaller density of states than NMs,
is negligibly weak compared with the intrinsic one.
 We also note that the spin injection problem seen in TDSs does not occur in NMs
because both the electrodes and conductor are metallic,
and the Hall angle is insensitive to $t^\prime$.

 The Hall conductance and Hall angle in the I-FM NM [Figs.~\ref{fig:VFM_NM}(a) and \ref{fig:VFM_NM}(c)] are much smaller than those in the V-FM NM [Figs.~\ref{fig:VFM_NM}(b) and \ref{fig:VFM_NM}(d)].
 This makes a good contrast with the ferromagnetic-electrodes-induced Hall effect in TDSs, 
where both the I-FM and V-FM show a large Hall angle (Fig.~\ref{fig:HallAngle}).
 The difference between the I-FM and V-FM NMs comes from the another consequence of the spin scattering: the spin diffusion.

\iftitle
 \subsection{Spin diffusion}
\else
 \textit{Spin diffusion.}
\fi
\label{sec:diffusion}

\begin{figure}[tbp]
 \centering
  \includegraphics[width=1\linewidth]{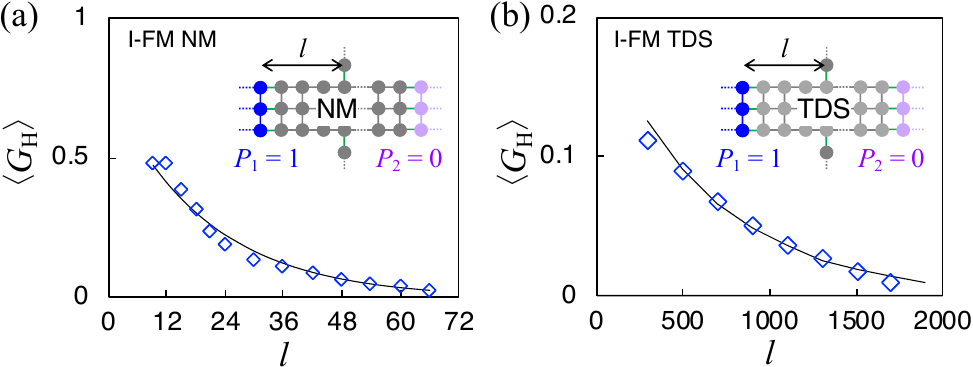}
 \vspace{-4mm}
\caption{
  Averaged Hall conductance $\braket{G\_H}$ as a function of the position of the voltage electrodes $l$ for (a) I-FM NM and (b) I-FM TDS
  with spin scatterers of $W_s=0.5$ and $\rho=5\%$.
  Only the source electrode is polarized ($P_1=1$ and $P_2=0$).
  The system size $L_x\times L_y \times L_z$ is (a) $72\times 12\times 12$ and (b) $2000 \times 20 \times 20$.
  The electrodes coupling is (a) $t^\prime=t$ and (b) $t^\prime=t/4$.
  The solid lines are the functions $c \exp(-l/\xi)$ with (a) $c=0.75$, $\xi=20$ and (b) $c=0.2$, $\xi=640$.
  Averaged over (a) 100,000 and (b) $800$ impurity realizations, and
  the error bars are less than the size of the symbols.
}
\label{fig:spinDecay}
\end{figure}

 The spin diffusion is one of the important problems in spintronics.
 Here we show that the spin-diffusion length in TDSs is extremely long compared with NMs.
 When only the source electrode is polarized (I-FM geometry with $P_1=1$ and $P_2=0$),
the injected spin current decays in the $x$ direction,
and the Hall effect decays in the same way.
 Therefore, the decay length $\xi$ of the Hall conductance $G\_H$ corresponds to the spin-diffusion length.
 Figure~\ref{fig:spinDecay} shows the averaged Hall conductance $\braket{G\_H}$ as a function of $l$, the distance from the FM source electrode to the voltage electrodes.
 In a long wire, the Hall conductance $\braket{G\_H}$ decays exponentially with increasing $l$.
 Note that this decay causes the small Hall conductance and small Hall angle in the I-FM NM [Figs.~\ref{fig:VFM_NM}(a) and \ref{fig:VFM_NM}(c)].
 The decay length $\xi$ in NMs is about $20$ sites (for $W_s=0.5$ and $\rho=5\%$), 
and it becomes shorter for a strong spin scattering.
 On the other hand, the decay length $\xi$ ($\simeq 6 \times 10^2$ sites, as a rough estimate) in TDSs is about $30$ times longer than in NMs with the same spin scattering strength.

\begin{figure}[tbp]
 \centering
  \includegraphics[width=1\linewidth]{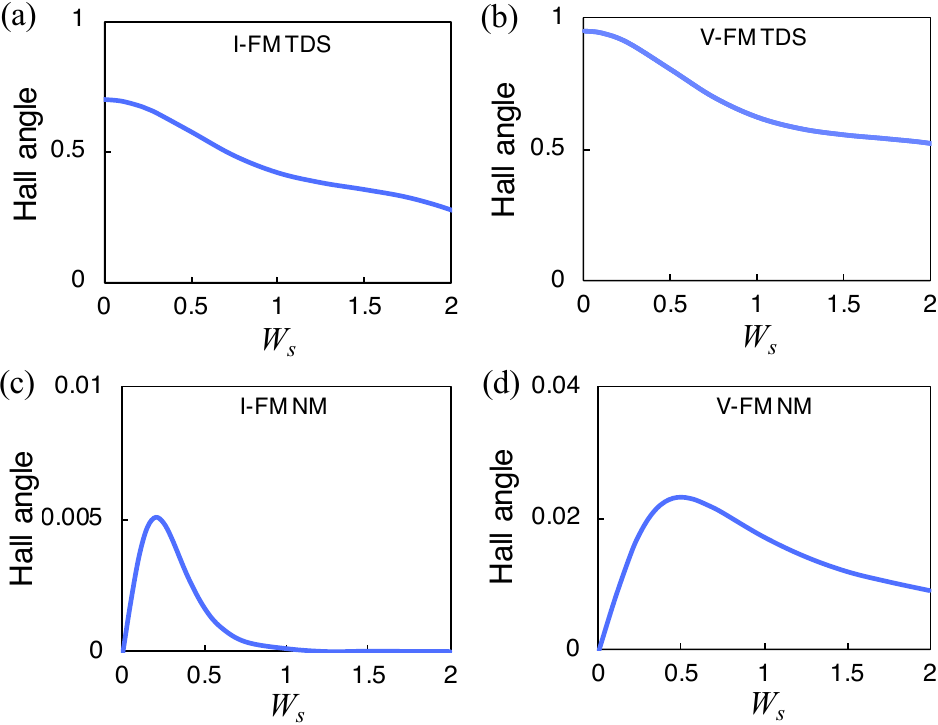}
 \vspace{-4mm}
\caption{
  Averaged Hall angle $\braket{R\_H}/\braket{R\_{SD}}$ for disordered 
  (a) I-FM TDS,
  (b) V-FM TDS,
  (c) I-FM NM, and
  (d) V-FM NM
  as functions of the scattering amplitude $W_s$, with $\rho=5\%$.
  The system size $L=20$, polarization $P_1=P_2=1$, and coupling $t^\prime=t/2$.
  Averaged over (a)(b) 1,000 and (c)(d) 20,000 impurity realizations so that the error bars are less than the width of the lines.
}
\label{fig:Wsdep}
\end{figure}

 Lastly, we mention an important feature of the ferromagnetic-electrodes-induced Hall effect in TDSs:
the stability against the spin scattering.
 Obviously, the ferromagnetic-electrodes-induced Hall effect in TDSs is robust against the spin-conserving impurity scattering,
since the effect originates from the helical surface states, i.e., the topological property.
 However, the spin scattering induces the backscattering into the helical surface states,
and it may destroy the ferromagnetic-electrodes-induced Hall effect.
 Figure~\ref{fig:Wsdep} shows the Hall angles for the I/V-FM TDSs and NMs as functions of the spin-scattering amplitude $W_s$ with the scatterer density $\rho=5\%$.
 Even though the impurity density is large,
the Hall angle in TDSs [Figs.~\ref{fig:Wsdep}(a) and \ref{fig:Wsdep}(b)] keeps more than one-half of the value in the clean limit for a strong spin-scattering $W_s\lesssim t=2$.
 This shows that the ferromagnetic-electrodes-induced Hall effect in TDSs are stable against the spin scattering.
 The Hall angle in NMs [Figs.~\ref{fig:Wsdep}(c) and \ref{fig:Wsdep}(d)], which comes from the extrinsic ferromagnetic-electrodes-induced Hall effect,
shows a maximum around $W_s=0.5$ ($\rho=5\%$ is also a maximum), and 
the maximum value is only a few percent of that in TDSs.
 The rapid decay of the Hall angle in the I-FM NM [Fig.~\ref{fig:Wsdep}(c)] corresponds to the decrease of spin-diffusion length.

\iftitle
 \section{Conclusions} \label{sec:conclusion}
\else
 \textit{Conclusion.}
\fi
 We have numerically studied the transport properties of the TDS with FM electrodes.
 We have found that an unconventional type of Hall effect is induced by the FM electrodes, with keeping the time-reversal symmetry of the TDS.
 The ferromagnetic-electrodes-induced Hall effect shows a large Hall angle $R\_H/R\_{SD}\simeq 1$ for ideal conditions,
which means that the Hall voltage $V\_H$ becomes as large as the source-drain voltage $V\_{SD}$,
like the anomalous Hall effect in the ideal magnetic WSM.
 In contrast to the anomalous Hall effect in magnetic WSMs, which is an inherent feature,
the strength of the ferromagnetic-electrodes-induced Hall effect can be externally controlled by the magnetization of the electrodes.
 In fact, the Hall response is maximized when the electrodes are magnetized in parallel and
vanishes when the electrodes are magnetized in antiparallel.
 Thus, the ferromagnetic-electrodes-induced Hall effect can be regarded as the Hall magnetoresistance effect.

 This Hall magnetoresistance is significant because the magnetoresistance ratio diverges in principle.
 The sensitivity to the magnetization of the electrodes implies that
we can detect the magnetization of a material attached to the TDS
(e.g., a memory cell), via the Hall response.
 Although the spintronics devices using the spin current often suffers from the problem of the efficiency of spin injection,
we have shown that the problem may not be serious
if the FM electrodes are the Hall voltage probes.
 We note that the TDS (\ce{Cd3As2}) with FM (Co) voltage electrodes is realized \cite{Lin20electric} recently. 
 We have also shown that the extrinsic ferromagnetic-electrodes-induced Hall effect coming from the spin scatterers is negligibly small compared with the intrinsic ferromagnetic-electrodes-induced Hall effect in TDSs,
and the Hall magnetoresistance effect is stable against impurities.

 Here, we comment on the condition for obtaining the large ferromagnetic-electrodes-induced Hall effect.
 The effect is not sensitive to the detail of the electrodes as long as the contact resistance is large enough, 
but highly polarized FM metals, such as half-metals, are preferable.
 In this paper, we have considered an ideal condition for the TDS where the transport due to the helical surface states are dominant.
 If a non-Dirac metallic band crosses the Fermi energy in addition to the helical surface bands,
the maximum value of the Hall angle decreases,
although it should be still finite as long as the helical surface states are not destroyed.
 The operating temperature of the device also depends on the energy range where the density of states is sufficiently small.
 Therefore, we expect a TDS with a small density of states near the Fermi energy to reproduce the diverging Hall magnetoresistance ratio from the ferromagnetic-electrodes-induced Hall effect.

 The large and robust ferromagnetic-electrodes-induced Hall effect is a characteristic of the helical surface/edge states.
 Thus, the results in this paper apply to any material with helical surface/edge states,
e.g., the two-dimensional quantum spin Hall insulators
(see Appendix~\ref{sec:2D})
such as \ce{HgTe} \cite{Bernevig06quantum} and transition metal dichalcogenides \cite{Qian14quantum}, 
thin films of topological insulators \cite{Kobayashi15dimensional}, or
higher-order topological insulators with helical hinge states \cite{Schindler18higher},
in addition to the TDSs \ce{Na3Bi} \cite{Wang12dirac, Liu14discovery} and \ce{Cd3As2} \cite{Wang13three, Neupane14observation}
(see Appendix~\ref{sec:material}).
 Recently, the kagome layered material Co$_{3-x}$Ni$_x$Sn$_2$S$_2$ \cite{Thakur20intrinsic,Shen20local} was suggested as a candidate for the TDS.
 The ferromagnetic-electrodes-induced Hall effect may be useful to identify the TDS state in such material by the transport measurement.


\begin{acknowledgments}
 We thank Yasufumi Araki and Junsaku Nitta for valuable discussions.
 This work was supported by the Japan Society for the Promotion of Science 
KAKENHI (Grant Nos.~%
JP19K14607 
and
JP20H01830) 
and
by CREST, Japan Science and Technology Agency (Grant No.~JPMJCR18T2).
\end{acknowledgments}

\begin{appendix}

\section{Quantum spin Hall systems} \label{sec:2D}
 We show that the ferromagnetic-electrodes-induced Hall effect is insensitive to the thickness 
as long as the helical surface/edge states are survived.
 Here we consider the single-layer limit of the model Eq.~\eqn{H_TDS},
which is
equivalent to the Bernevig-Huges-Zhang model \cite{Bernevig06quantum} describing
the quantum spin Hall (QSH) insulators.
 The model shows the nontrivial ($\mathcal{N}_s=+1$) phase for $0<m_0/m_2<2$,
and we set $m_0/m_2=1$.

\begin{figure}[tbp]
 \centering
  \includegraphics[width=1.\linewidth]{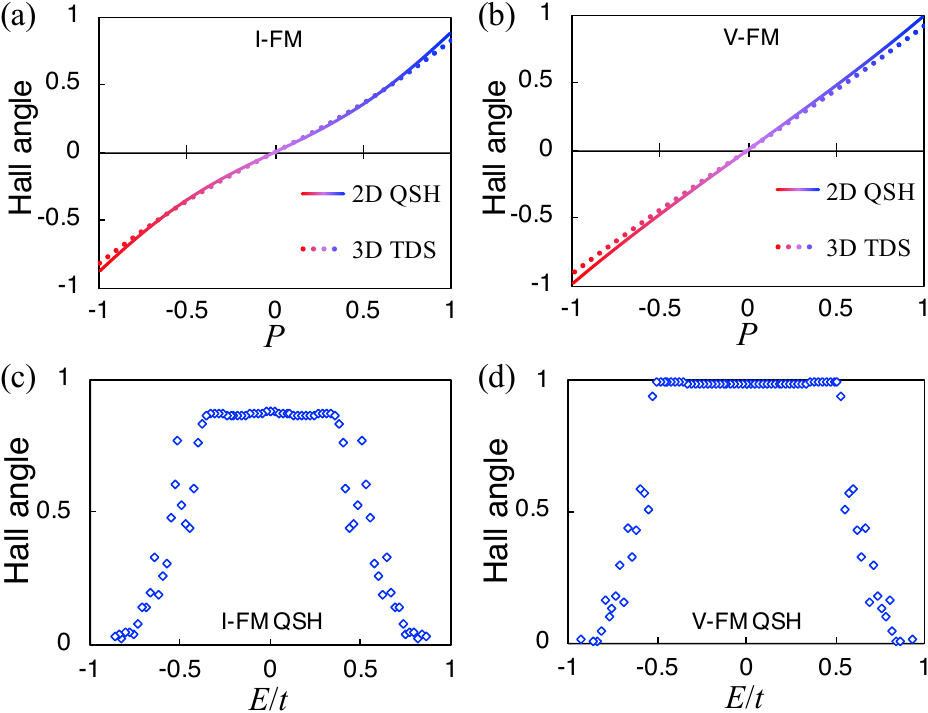}
 \vspace{-5mm}
\caption{
  Hall angle $R\_H/R\_{SD}$ for 2D QSH insulators as a function of $P$ for the 
  (a) I-FM and (b) V-FM geometries.
  The Hall angle in 2D (solid, $m_0/m_2=1$, $L_z=1$) is slightly larger than in 3D (dashed, $m_0/m_2=2$, $L_z=20$).
  The energy dependence of the Hall angle for the 
  (c) I-FM and (d) V-FM geometries.
  The size $L_x=120, L_y=20$ and the coupling $t^\prime=t/4$.
}
\label{fig:2D3D}
\end{figure}

 Figure~\ref{fig:2D3D} shows the robustness of the effect against the change of the thickness. 
 This indicates that the same discussions as the main text apply to $s_z$-conserved QSH systems (such as HgTe) or thin films of TDSs.
 Since the spin injection problem arises in the same manner as 3D TDSs,
the advantage of the V-FM also holds in the 2D limit.
 Although the realization of perpendicular magnetizations in thin films of electrodes may be challenging,
the ferromagnetic electrodes induced Hall effect in thin films is more efficient;
the Hall angle is large inside the bulk energy gap $|E|<t/2$ [see Fig.~\ref{fig:2D3D}(c) and \ref{fig:2D3D}(d)].

 When the scattering between the edge channels is negligible, i.e., the top and bottom edges are sufficiently separated,
the conductances/resistances will be quantized.
 By assuming the ballistic transport via the edge states, 
we can analytically obtain the transport properties from the Landauer-B\"{u}ttiker formalism.
 The quantized values of the four-terminal conductances/resistances with paramagnetic or half-metallic electrodes as listed in Table~\ref{tab:GR2D}.

\begin{table}[tbp]
   \centering
   \topcaption{
Conductances and resistances for four-terminal ideal QSH systems with various electrodes in units of $e^2/h$ and $h/e^2$, respectively. The system shows QSH effect for normal (paramagnetic) electrodes. When all the electrodes are ferromagnetic, the system is effectively equivalent to a quantum anomalous Hall system.
}
   \begin{tabular}{@{} ccccc @{}}
      \toprule[0.75pt]
       & \multicolumn{4}{c}{Electrodes} \\
      \cmidrule[0.4pt](r){2-5}
           & Para & All-FM & I-FM & V-FM \\
      \midrule[0.4pt]
      $G\_{SD}$  & 1 & 1   & 1   & 2   \\
      $G\_{H}$   & 0 & 1/2 & 1/2 & 1   \\
      $R\_{SD}$  & 1 & 1   & 1   & 1/2 \\
      $R\_{H}$   & 0 & 1   & 1   & 1/2 \\
      \bottomrule[0.75pt]
   \end{tabular}
   \label{tab:GR2D}
\end{table}

\section{\ce{Na3Bi} and \ce{Cd3As2}} \label{sec:material}
 We illustrate the results in the main text performed on a minimal TDS model apply to realistic TDSs.
 For specific examples of the TDS, 
we employ the effective model up to the quadratic terms \cite{Wang12dirac,Cano17chiral,Ominato19spin},
\begin{align}
 H &=  \sum_{\bf r}
         \[ {it \over 2} 
           \(
              \Ket{{\bf r}+{\bf e}_x}
               \tau_x \sigma_z
              \Bra{\bf r}
            + \Ket{{\bf r}+{\bf e}_y}
               \tau_y \sigma_0
              \Bra{\bf r}
           \)
           + \textrm{H.c.}
         \] \nonumber \\
   &+  \sum_{\bf r}
         \[ \Ket{{\bf r}+{\bf e}_z}
           \(
              -{m_{1} \over 2} \tau_z\sigma_0
              -{C_{1} \over 2} \tau_0\sigma_0
           \)
           \Bra{\bf r}  + \textrm{H.c.}
         \]  \nonumber \\
   &+  \sum_{\bf r}
        \sum_{\mu=x,y}
         \[ \Ket{{\bf r}+{\bf e}_\mu}\!
           \(
              -{m_{2} \over 2} \tau_z\sigma_0
              -{C_{2} \over 2} \tau_0\sigma_0
           \)\!
           \Bra{\bf r}  + \textrm{H.c.}
         \]   \nonumber \\
   & + \sum_{\bf r} \Ket{\bf r}
        \( m_0 \tau_z\sigma_0
          +C_0 \tau_0\sigma_0
        \) \Bra{\bf r},
 \label{eqn:H_FP}
\end{align}
with material parameters of \ce{Na3Bi} and \ce{Cd3As2} from first principle calculations \cite{Wang12dirac,Cano17chiral} (see Table~\ref{tab:param}).

\begin{figure}[tbp]
 \centering
  \includegraphics[width=1.\linewidth]{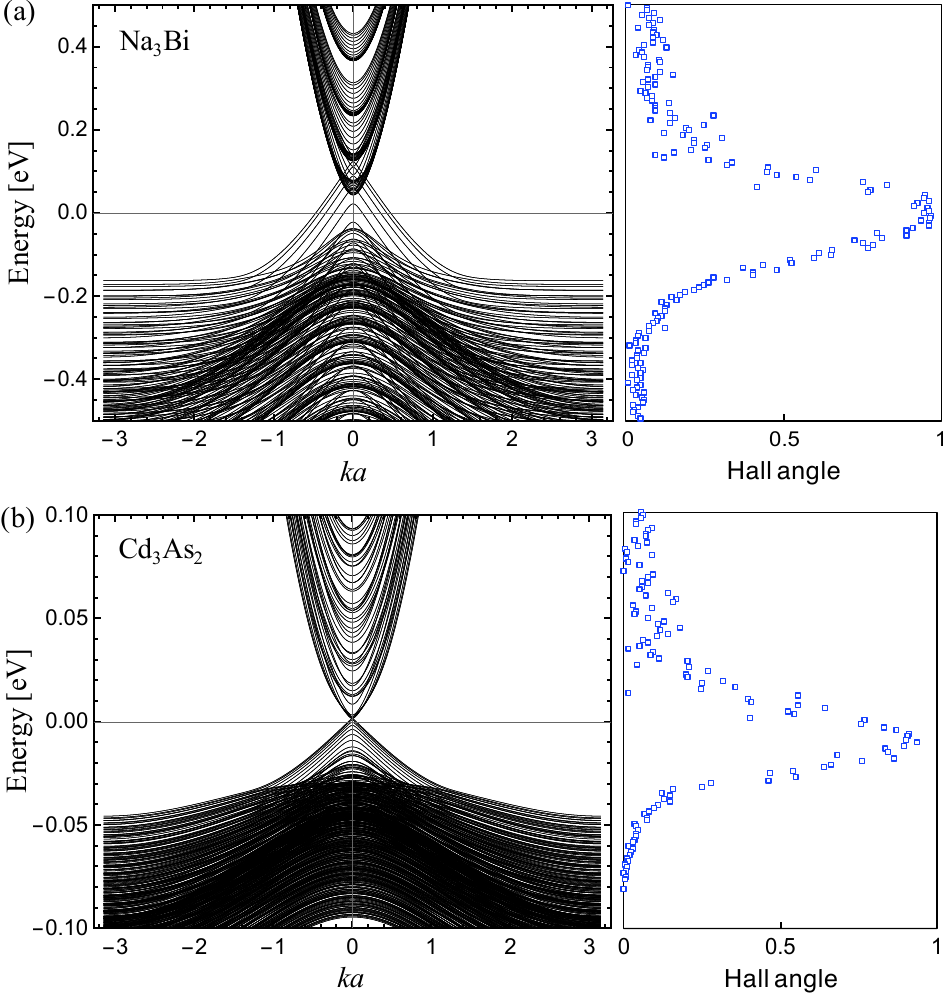}
 \vspace{-3mm}
\caption{
  Band structure of the nanowire (left panels) and V-FM Hall angle (right panels) as a function of energy in (a) \ce{Na3Bi} and (b) \ce{Cd3As2}.
  The system size $L=20$, and the electrodes are half-metallic $P=1$ with coupling $t^\prime=t$.
}
\label{fig:material}
\end{figure}

 The band structures for $L_y=L_z=20$ and the corresponding Hall angles are shown in Fig.~\ref{fig:material}.
 The Hall angle is large in the range where the helical surface states arise in the band structure
and vanishes where the density of states is large.
 In both \ce{Na3Bi} and \ce{Cd3As2}, the maximum value of the Hall angle is as large as the ideal value $1$ around the Dirac point $E=0$.
 For $E=0$, we have confirmed the dependence of the Hall angle on the polarization of electrodes and the effect of the spin injection are the same as those in the minimal model Eq.~\eqn{H_TDS}.

\begin{table}[htbp]
   \centering
   \topcaption{
Material parameters. Note that the signs of $C_0,C_1$, and $C_2$ for \ce{Na3Bi} is reversed from those in Ref.~\cite{Wang12dirac} so as to reproduce the density of states around the Dirac point.
}
   \begin{tabular}{@{} c@{\hspace{5mm}}rl@{\hspace{5mm}}rl @{}}
      \toprule[0.75pt]
           & \multicolumn{2}{l}{\ \ \ce{Na3Bi}} & \multicolumn{2}{l}{\ \ \ce{Cd3As2}} \\
      \midrule[0.4pt]
      $C_0$       & $ 1.183$ &eV  & $ 0.306$ &eV \\
      $C_1$       & $-0.118$ &eV  & $ 0.033$ &eV \\
      $C_2$       & $ 0.654$ &eV  & $ 0.144$ &eV \\
      $m_0$       & $ 1.754$ &eV  & $ 0.376$ &eV \\
      $m_1$       & $-0.228$ &eV  & $-0.058$ &eV \\
      $m_2$       & $-0.806$ &eV  & $-0.169$ &eV \\
      $t$         & $ 0.485$ &eV  & $ 0.070$ &eV \\
      $a$         & $ 5.07$  &\AA & $12.64$ &\AA \\
      $c$         & $ 9.66$  &\AA & $25.43$ &\AA \\
      \bottomrule[0.75pt]
   \end{tabular}
   \label{tab:param}
\end{table}

\end{appendix}

\bibliography{FMlead}

\end{document}